\documentclass[a4paper,fontsize=9pt,twocolumn]{scrartcl}
\usepackage[pdftex]{graphicx}
\usepackage[cmex10]{amsmath}
\usepackage{array}
\usepackage{graphicx}
\usepackage{amssymb}
\usepackage{amsthm}
\usepackage{mathptmx}
\usepackage[T1]{fontenc}
\usepackage{bbding}
\usepackage[all]{xy}
\usepackage{subfigure}

\usepackage{dsfont}
\usepackage{array}
\usepackage{rotating}
\usepackage{amsmath}
\usepackage{rotating} 																	
\usepackage{url}
\usepackage{setspace}
\usepackage{verbatim}

\DeclareMathAlphabet{\mathcal}{OMS}{cmsy}{m}{n}
\newcommand{\oss}{\mathcal{O}}
\newcommand{\kmat}{\mathcal{K}}
\newcommand{\ns}{\mathcal{N}}
\newcommand{\I}{\mathrm{I^m}} 													
\newcommand{\V}{\mathrm{V^m}} 													
\newcommand{\Ie}{\mathrm{I}} 														
\newcommand{\at}{\widetilde{a}}%
\newcommand{\bt}{\widetilde{b}}%
\newcommand{\ct}{\widetilde{c}}%
\newcommand{\dt}{\widetilde{d}}%

\newtheorem{thm}{Theorem}
\newtheorem{lem}{Lemma}
\newtheorem{mydef}{Definition}
\newtheorem{cor}{Corollary}
\newtheorem{rem}{Remark}

\usepackage[colorinlistoftodos,textsize=tiny,textwidth=1.4cm]{todonotes}

\usepackage{authblk}
\usepackage[font={small,it}]{caption}

\begin{document}
\title{Biological mechanism and identifiability of a class of stationary conductance model for
Voltage-gated Ion channels
}

%

\author[1]{Febe~Francis}
\author[2]{M{\'i}riam~R.~Garc{\'i}a \footnote{E-mail: miriamr@iim.csic.es}}
\author[1]{Oliver~Mason}
\author[3]{Richard~H.~Middleton}
\affil[1]{\small{Hamilton Institute, National University of Ireland, Maynooth,Co. Kildare, Ireland.}}
\affil[2]{Bioprocess Engineering Group, IIM-CSIC, Vigo, Spain}
\affil[3]{Complex Dynamic Systems \& Control, The University of Newcastle, Australia.}

\maketitle

\begin{abstract}

The physiology of voltage gated ion channels is complex and insights into their gating mechanism is incomplete.
Their function is best represented by Markov models with relatively large number of distinct states that are connected by thermodynamically feasible transitions. On the other hand, popular models such as the one of Hodgkin and Huxley have empirical assumptions that are generally unrealistic. Experimental protocols often dictate the number of states in proposed Markov models, thus creating disagreements between various observations on the same channel. Here we aim to propose a limit to the minimum number of states required to model ion channels by employing a paradigm to define stationary conductance in a class of ion-channels. A simple expression is  generated using concepts in elementary thermodynamics applied to protein conformational transitions. Further, it matches well many published channel current-voltage characteristics and parameters of the model are found to be identifiable and easily determined from usual experimental protocols.

\end{abstract}


\section{Introduction}
\label{intro}

The electrical activity of a living system is a dynamic function of the ionic transport across biological membranes. Ion channels are 
pore-forming protein ensembles that are responsible for the task of regulating ion flows. Gating arises as conformational changes in the 
proteins that comprise the channel.
These conformational changes are driven by changes in the electric field or by molecules (ligands) that bind to them. For this reason, ion-channels are often classified into voltage-gated and ligand-gated categories.

Voltage-gated ion channels have charged domains that make their structure sensitive to variations in the external electric field. For a 
particular range of membrane potentials, they adopt a conformation with a central hole: forming a channel for the free movement of ions. 
Such an `open' state is further defined by certain `selectivity filters' (often amino acids) that would render specificity for the protein 
\cite{Beyl2007}. At other membrane potentials, the flow of ionic current is blocked as a result of the `closed' or `inactive' 
conformations that the protein adopts. A channel protein can thus adopt various conformational states with varying degrees of conductance, 
and they can spontaneously switch between these states. The steady distribution and dynamics of such switches is central to any study that 
involves an ion channel.

Equation based kinetic models are useful to interpret the behaviour of a channel in a given situation. Starting with the model of Hodgkin 
and Huxley \cite{HODGKIN1952}, several researchers have developed theoretical frameworks that partially explain observations made on 
channel activity. 
The Hodgkin-Huxley formalism relies on an underlying model of average channel conductance and their equations describe the changes of ionic 
permeability with membrane potential.  The model makes use of hypothetical gating particles to bring about the channel's function, by 
forcing their motion with respect to the electric field across the membrane. Although this model has been used in many instances, 
hypothetical gating particles do not appear to be consistent with underlying molecular mechanisms. 

Further developments in the study of ion-channels made an attempt to give a
mechanistic description of the gating phenomena. Such models consist of nonlinear ordinary differential equations (ODEs), including a 
current balance equation and the dynamics of conformational transitions incorporated as a `gating variable' that corresponds to the state 
of the ion channels. Discrete-state Markov models have been  used to describe the different states of the ion channel \cite{Kienker1989} 
and have the advantage of providing a mechanistic description for the otherwise abstract Hodgkin-Huxley formalism based models. 

Markov chain models are developed on the assumption that ion channels exist in a finite number of significant energy states, with 
time-homogeneous rates of transition between them. The model consists of a topology of allowed transitions between these states, together 
with the rates for these transitions. Fitting single-channel recording data with a Markovian kinetic scheme has been standard in 
neurophysiology for quite some time \cite{colquhoun-hawkes:81,Kienker1989,milescu-akk-sachs:05, Sansom19891229}. However, for good 
agreement with experimental data, frequently the number of closed states needed varies with the experimental protocol.  A significant 
benefit of Markov models compared to the Hodgkin-Huxley formalism is that the large degree of freedom in the model stucture allows it to 
fit more closely with experimental observations \cite{Fink2009}. Thermodynamic models are a two state Markovian description of channel 
flipping, the rate kinetics of which are described by concepts in thermodynamics \cite{Destexhe2000, Ozer2004}.

Fractal models \cite{LiebovitchLS1987, Liebo89} of ion channel gating provide a different description of the underlying mechanism compared 
to Markov models. Such models are characterised by equations having continuous rather than discrete states. The Diffusion models introduced 
by Millhauser et al. \cite{Millhauser}, justify Fractal models at a microscopic level. Statistical analysis, however has often favored 
Markov 
models over Fractal models \cite{Sansom19891229,Nek2007}.

A major hurdle in modelling ion-channel gating using a Markov-jump scheme is in the determination of an appropriate number of closed 
states. 
As the topology space expands with the number of states, generating appropriate kinetic schemes can give rise to ambiguity because it may 
be 
possible to come up with multiple schemes that are consistent with a given set of data. Moreover, numerical simulation of such models
becomes time-consuming. Thus the model has limitations from the point of view of parameter estimation and further in its use for 
multi-cellular simulations \cite{Fink2009}.  The increasing complexity of the topology generates a need for model 
reduction. Kienker \cite{Kienker1989} talks about the existence of equivalence in topologies of models that are identifiable within the 
same data 
set. This would imply that models with a larger
number of states are reducible. Keener \cite{Keener2009} illustrates the possibility of reducing the complexity to stable invariant 
manifolds. 
This approach would reduce the dimension of the system without suffering from large approximation errors. Furthermore, the time scale of 
the Markovian transitions are much faster than the main time scales involved in the aggregate cell voltage and ion behaviour 
\cite{DavidJ.Aidley1996, hille}, which in turn offers options for model reduction.

Another difficulty in the acceptability of ion channel models is related to parameter identifiability. A model is said to have structurally 
unidentifiable parameters when multiple parameters are equally powerful in explaining observed data in noise-free perfect experiments 
\cite{bellman1970structural,walter1997,balsa2010dynamic}. 
Since it is well-known that a large segment of ion-channel models in the literature lack parameter identifiability in noise experiments 
\cite{Fink2009}, fundamental studies of structural identifiability are scarce due to the difficulty of solving the associated symbolic set 
of equations \cite{csercsik-szederkenyi-hangos:2010,csercsik2012identifiability}.

It should be also noted that in usual experimental protocols optimised to improve the signal-to-noise ratio 
\cite{beaumont1993interpretation,willms1999improved}, provide two sets of separated data: one used to characterise the steady-state, and 
the other with time constants, when dynamics are not sufficiently fast to be disregarded.  With this in mind, Hodgkin-Huxley models employ 
empirical expressions for both steps that are not always realistic \cite{willms1999improved}.  On the other hand, estimation in models 
based on Markov chains often make use of other types of experimental protocols, and do not exploit the large range of data available using 
the standard protocols.

In this article, we describe a voltage-gated ion-channel model of stationary conductance with three main characteristics.  To begin with, 
the model has a clear mechanistic motivation based on the underlying thermodynamics.  In addition, the model is relatively simple to 
implement, with a small number of easily identifiable parameters.  Finally, we test the model using experimental data from patch clamp and 
similar studies reported in the literature.  Our model can make use of classical experimental protocols and provide a mechanistic 
formulation for the calibration of steady-state characteristics in the Hodgkin-Huxley setting and, by identifying the form of the 
stationary 
distribution, constrain the estimation process in Markov processes, where the study of global structural identifiability is still an open 
problem.  In addition, the simple model proposed here can be applied to study the gating of fast ion channels, such as fast persistent and 
fast activated Na$^+$ currents or transient activated K$^+$ 
current \cite{Izhikevich2007dynamical}.

\section{Background}\label{sec:background}

\subsection{Stable conformations of ion channels }

The probability of a protein molecule adapting a particular conformation in space is largely influenced by the electrical force field 
surrounding it. When subject to an electric field, charged groups within a protein will experience a force and may attain a new 
electrostatic equilibrium by incorporating angular changes in the dipoles associated with the peptide bond \cite{Manna2007}. If the 
thermodynamic kinetic energy is large enough to overcome the energy barrier, the protein takes
up a new conformational state. Proteins can hence take up a large number of conformational states separated by small energy barriers. 
Despite the continuum of intermediate states, this dynamical system would typically have only a few stable equilibria \cite{Keener2009} 
and hence a limited number of experimentally observable states.

\subsection{Transition velocities }
\label{ss:tv}

The rate at which the protein switches between such stable states is mainly determined by the driving forces that help in overcoming the 
thermodynamic energy barrier. However, there are some conformational changes that are not regulated by either electric field or ligands. 
They may be considered to have a constant energy barrier in the given environment, and hence may be thought to have a uniform rate. 
Classical thermodynamics identifies the rate of transition between two reaction states based on the free energy barrier between them as

\begin{equation}
k=k_{0} e^{-{(\triangle G)}/{RT}},\end{equation}
where $k$ is the rate of transition between the two states, $k_{0}$ is a  constant, \textit{\ensuremath{\Delta}G} is the free energy 
barrier 
between the two states, $R$ is the universal gas constant and $T$ the absolute temperature.

\begin{figure}[!ht]
\includegraphics[width=\linewidth]{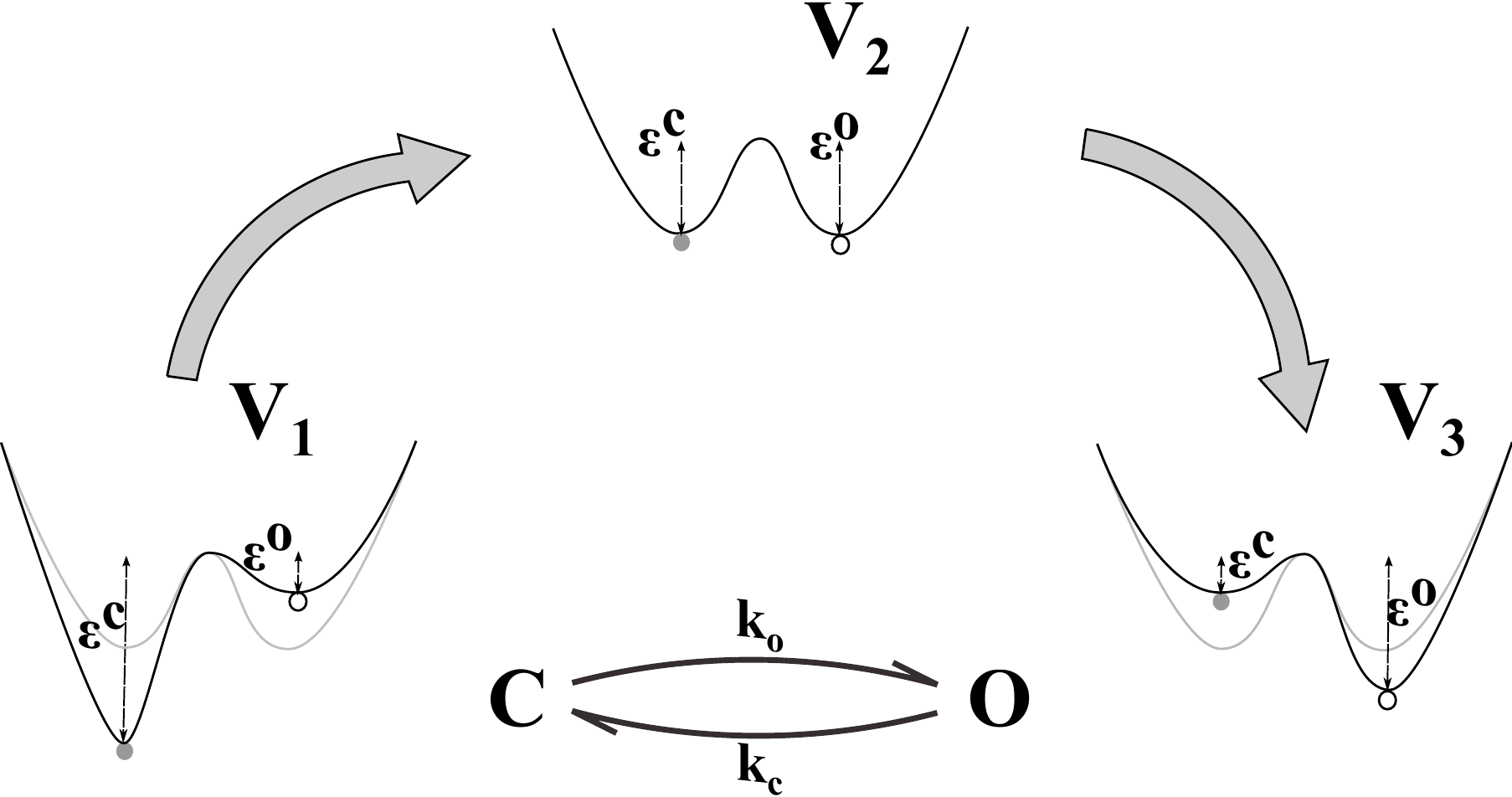}
\caption{ Energy profile for the transitions between open and closed states in an ion channel, modulated by a change in membrane potential: 
a change in membrane potential brings in protein conformational changes and thereby free energies of protein conformation. The rates of 
channel closing and opening, $k_c$ and $k_o$, are a function of the free energy barriers ($\epsilon^c$ and $\epsilon^o$) defined by 
structural configurations of the respective states. The figure describes how the  open probability of a channel is enhanced by a change in 
the ambient voltage from $V_1$ to $V_3$ by virtue of its conformational energy. }
\label{fig:gibbs}
\end{figure}

More generally, the free energy barrier may be dependent on the electric field, that is, the membrane potential $V$ (figure 
\ref{fig:gibbs}). 
In this case we have 
 \begin{equation}
k(V)=k_{0}e^{-{\ensuremath{\Delta} G(V)}/{RT}}\end{equation}
Here \textit{\ensuremath{\Delta}G(V)} is the free energy barrier between the two states defined by the voltage \cite{milescu-akk-sachs:05}.

In the case of voltage based transition, the activation energy can be expressed 
in a general form by using a Taylor series expansion \cite{Destexhe2000, Ozer2004}, as follows:
\[
\ensuremath{\Delta}G(V) ={a+bV+cV^{2}+\cdots}
\]
and the rate of transition \textit{k(V)} may be written as,
\begin{equation}
k(V)=k_{0}e^{-{(a+bV+cV^{2}+\cdots)}/{RT}}\label{eq:gibbsvolt}
\end{equation}
Here $a$ corresponds to the free energy independent of the electric field and $bV$ corresponds to interactions between the electric field 
and isolated charges and rigid dipoles on the protein. The higher order terms correspond to the influence of polarization and deformation 
within the protein structure as well as mechanical constraints. These effects are usually negligible as the trans-membrane voltage 
variations are generally small \cite{Destexhe2000}. 

In what follows, we are mainly interested in models where the free energy barrier is linear in the membrane potential. The effect of 
temperature variations may be neglected for the system under consideration, as long as the physiological environment remains unaffected. In 
this case, the following trivial lemma, helps us derive a simple expression for voltage regulated conformational transitions.

\begin{lem}
\label{lemone}
Consider a set $\{\ensuremath{\Delta} G_1(V), \dots , \ensuremath{\Delta} G_M(V)\}$ of activation energies, where each $\ensuremath{\Delta 
G_i}(V)$ is affine in V.  Denote the corresponding transition rates by $\{k_1, \ldots, k_M\}$.  Then any expression of the form
$\prod_{p=1}^P {\left(\frac{k_{i_p}(V)}
{k_{j_p}(V)}\right)}$ may be expressed as $e^{\left[(V-V_{h})s\right]}$
\end{lem}

\subsection{Markov Models for Conformation transitions}

The series of molecular changes associated with the opening of an ion channel is often described using Markov models with discrete states 
\cite{Kienker1989, Sansom19891229, Keener2009}.   To begin, consider the simple case of a transition between an open and closed state; let 
$O$ and $C$ represent the probabilities that the molecule is in the corresponding open and closed state at a
given time. Since the equation governing changes in probabilities for a single molecule has a form similar to the rate equation for a large 
number of molecules, the transition may be represented by a kinetic scheme, as follows:

\[
\xymatrix{C\ar@<.5ex>[r]^{k_{OC}} & O\ar@<.5ex>[l]^{k_{CO}}},
\]
where $k_{ij}$ represents the rate of transition from state $j$ to state $i$. The above kinetic scheme has a transition intensity matrix 
$\kmat$ \[ \left[\begin{array}{cc} -k_{CO} & k_{OC}\\ k_{CO} & 
-k_{OC}\end{array}\right] \] 
and the steady state probability for the channel to be in an open state leads 
\[ \oss =\frac{1}{1+\frac{k_{OC}}{k_{CO}}} .\]
Using Lemma (\ref{lemone}) we obtain,
\begin{equation}
\oss =\frac{1}{1+e^{\left[(V-V_{h})s\right]}}\label{eq:1}
\end{equation}
The expression is the modified Boltzmann's expression used in modeling ion-channel gating \cite{Beyl2007, Huang, Xu2001}. This sigmoidal 
function of voltage is symmetric about the half-activation voltage, $V_h$. However, experimental data frequently show one or both of: (i) 
activation and inactivation behaviour (ii) asymmetric behaviour, and hence it is hard to fit data with a single Boltzmann function. This 
implies that the two-state Markov chain with linear energy barriers does not ably model experimental observations. 

Three possible ways to resolve this are (i) model the system as having more than two macro-states (stable conformational states), (ii) 
model 
the system as having transition rates that change with the dwell times, leading to a fractal model; or, (iii) use non-linear terms in the 
energy expression. 

We focus on the first option here for the following reasons.  The second option (ii) would give rise to time-inhomogeneous Markov chains, 
which are typically very difficult to analyse. Moreover, the identification task becomes less tractable in this case.  
As mentioned previously, the last option (iii) includes higher order terms that are usually neglected. Also, as we shall see later in 
section (\ref{sec:nl}) the use of higher order non-linear energy models does not appear to result in simpler models, nor does it give more 
accurate fits to experimental data.

In fact, usually models developed on the basis of Markovian dynamics have more than three states [option (i)] for a good  agreement with 
experimental observations \cite{Kienker1989, milescu-akk-sachs:05}. In such models the rate constant is often defined with an exponential 
or a Boltzmann function in voltage, both obtained from a linear approximation of equation \eqref{eq:gibbsvolt}. 

We will now analyse how best we can extend the two-state model by option (i) to have a network with limited and identifiable macro-states.

\section{A Multiple Conformation Extension of the `Modified Boltzmann Function'
for stationary conductance}

\subsection{The Master Equation and Transition Rates}
\label{secME}
Here we consider the transition network of the ion-channel as a system with a set of $\ns$ stable states, marked $i = 0,1,...,n$; with 
$n=\ns-1$ closed or inactive states and $0$ being the open state. 
In the following, we consider models with a single open state, as this reflects the molecular structure of many ion channels as detailed in 
experiments and is not a generalization \cite{Liebo89,Silva2009} 
 If $S_{i}$ denotes the probability for the protein to be in state $i$ at any time $t$, the system obeys the \textsl{master equation} 
\begin{equation}
\dot{S}=\kmat S,
\label{eq:mastereqn}
\end{equation}
where $S = (\oss, S_{1},S_{2},...,S_{n})$  and $\kmat \in \mathbb{R}^{\ns 
\times \ns}$ is a transition matrix with $k_{ij} \geq 0$ giving the rate of 
transit from state $j$ to state $i$. The diagonal elements satisfy, $k_{ii}= 
-\sum_{i \neq j} k_{ij}$, to ensure the system evolves on the probability simplex.  The entry 
$k_{ij}\neq 0$ if and only if there is a transition from state $j$ to state $i$. 
The stationary probability distribution $S$ satisfies $\kmat S = 0$, $\mathds{1}_{\ns}^{T}S = 1$, where $\mathds{1}_{\ns}$ 
is the column vector of size $\ns$ with all entries equal to one. 

As is standard, we associated a directed graph with the Markov process described by (\ref{eq:mastereqn}) consisting of the 
nodes $\{0, 1, \ldots, n\}$ with an edge from state $j$ to state $i$ ($i \neq j$) if and only if $k_{ij} \neq 0$.  

In the following subsection, we derive explicit formulae for the form of the stationary vector in some simple cases; emphasising
that in these cases the open state probability $\oss$ takes a particularly simple form.  We also describe a general condition on the 
structure
of the graph associated with the Markov process that is sufficient for this simple form to hold.

\subsection{Examples with similar form of solution}

\subsubsection{Solution for a special case: A three state linear system 
with reversible transitions}

A three state transition diagram for channel opening is given below
\[
\xymatrix{C_{1}\ar@<.5ex>[r]^{k_{21}} & C_{2}\ar@<.5ex>[l]^{k_{12}}
\ar@<.5ex>[r]^{k_{02}} & O\ar@<.5ex>[l]^{k_{20}}}.
\]
Here $O$ represents an open state and $C_1$ and $C_2$ represent closed or inactive conformations. The steady-state probability for the 
channel to be in the open state can be calculated as
\[
\oss=\frac{1}{\left(1+\frac{k_{12}}{k_{21}}+\frac{k_{12}}{k_{21}}
\frac{k_{20}}{k_{02}}\right)}. 
\]
By using lemma \ref{lemone}, there exist $V_{h_1}, s_1, V_{h_2}, s_2$ such that
\begin{equation}
\oss=\frac{1}{1+e^{\left[(V-V_{h_1})s_1\right]}+e^{\left[(V-V_{h_2})s_2\right]}}
\label{eq:middleton1}.
\end{equation}

In the case of a slightly different topology,
\[
\xymatrix{C_{1}\ar@<.5ex>[r]^{k_{01}}&O\ar@<.5ex>[l]^{k_{10}}\ar@<.5ex>[r]^{k_{20}} 
& C_{2}\ar@<.5ex>[l]^{k_{02}}}
\]
the probability of the channel to be in an open conformation would be

\[ \oss=\frac{1}{\left(1+\frac{k_{10}}{k_{01}}+\frac{k_{20}}{k_{02}}\right)} \]
and once again, using lemma \ref{lemone}, the stationary probability of being 
open can be represented in the form \eqref{eq:middleton1}.

\subsubsection{Linear Networks}

This scheme can be generalized for $n$ macrostates depending on the position of the open state in the entire topology. A topology involving 
the open state on the network extremum,
\begin{equation*}
\xymatrix{O\ar@<.5ex>[r]^{k_{10}} &
C_{1}\ar@<.5ex>[l]^{k_{01}}\ar@<.5ex>[r]^{k_{21}} & 
C_{2}\ar@<.5ex>[l]^{k_{12}} \ar@{.}[r] &
C_{i} \ar@{.}[r] &
C_{n-1}\ar@<.5ex>[r]^{k_{n,n-1}} & 
C_{n}\ar@<.5ex>[l]^{k_{n-1,n}}}
\end{equation*}
would have an open state probability
\begin{equation*}
\oss=\frac{1}{\left(1+\displaystyle\sum_{j=1}^{n}\prod_{i=1}^{j}
\frac{k_{i,i-1}}{k_{i-1,i}}\right)}
\end{equation*}
and a topology
\begin{equation*}
\xymatrix{C_{1}\ar@<.5ex>[r]^{k_{21}} &
C_{2}\ar@<.5ex>[l]^{k_{12}}\ar@<.5ex>[r]^{k_{32}} &
\cdots C_{p-1}\ar@<.5ex>[l]^{k_{23}} \ar@<.5ex>[r]^{k_{0,p-1}} &
O\ar@<.5ex>[l]^{k_{p-1,0}}\ar@<.5ex>[r]^{k_{p+1,0}} &
\cdots C_{n-1}\ar@<.5ex>[l]^{k_{0,p+1}} \ar@<.5ex>[r]^{k_{n,n-1}} & 
C_{n}\ar@<.5ex>[l]^{k_{n-1,n}}}
\end{equation*}
would yield
\begin{equation*}
\oss =\frac{1}{\left(1+\displaystyle\sum_{j=1}^{p-1}\prod_{i=1}^{j}\frac{k_{i,i+1}}{k_{i+1,i}}+
\sum_{j=p}^{n}\prod_{i=1}^{j}\frac{k_{i+1,i}}{k_{i,i+1}}\right)}.
\end{equation*}
In either case, with respect to the argument in lemma (\ref{lemone}), the open state probability can be reduced and in general, a system 
with $N$ macro-states related in order of their conformational transitions would have an open state probability
\begin{equation*}
\oss =\frac{1}{1+\displaystyle\sum_{i=1}^{n}e^{(V-V_{h,i})s_{i}}},
\end{equation*}
where \textit{$n=\ns-1$}, is the number of transitions in the linear network.

\begin{rem}
\label{rem:absorb}
For the networks considered so far, rendering one of the transitions 
irreversible, would make the network absorbing in nature. In other words, the 
system would reach and never leave a fixed conformation. Since such a possibility 
is not physically reasonable for ion-channels under consideration, this case is 
not considered further. 
\end{rem}

\subsubsection{Other networks}
More generally, consider a transition network in which a unique simple 
path exists from every stable state to the open state; so in the directed graph
associated with the matrix $\kmat$, there is a unique path from every $j \neq 0$ to the node
$0$.  We also assume that the matrix $\kmat$ is irreducible \cite{HJ1}.

The celebrated Markov Chain Tree Theorem \cite{LR86} allows us to characterise the form of the 
steady state probability of the open state in this case.  This result is usually stated for column stochastic matrices 
or discrete Markov chains; however, it is trivial to see that an exact analogue also holds for continuous chains with
matrices of the form $\kmat$.  We state a restricted version of this result below but first introduce some notation.  

For the directed graph $G$ associated with $\kmat$, a rooted spanning tree $T_i$ at $i \in \{0, \ldots, n\}$ consists of
the vertices $\{0, 1, \ldots, n\}$ and has the following properties: (i) $T_i$ is acyclic; (ii) for every $j \neq i$, there exists exactly 
one
outgoing edge from $j$; (iii) there exist no edge outgoing from $i$.  We denote by $w(T_i)$ the weight of $T_i$, which is given by the 
product of
the entries of $\kmat$ corresponding to the edges in $T_i$.  For $0 \leq i \leq n$, let $\mathcal{T}_i$ denote the set of all directed 
spanning trees rooted at $i$, and define $w_i = \sum_{\mathcal{T}_i} w(T_i)$.   Note that  each $w_i$ will be a sum of terms of the form 
\begin{equation}
\label{eq:prodterm}
k_{i_1j_1}k_{i_2j_2} \cdots k_{i_nj_n}
\end{equation}

The following result is now a simple re-wording of the Markov Chain Tree Theorem as presented in \cite{LR86} and elsewhere.
\begin{thm}
\label{thm:MCT} Assume the matrix $\kmat$ is irreducible.  The unique stationary probability vector associated with $\kmat$, $\pi$ is given 
by
$$\pi_i = \frac{w_i}{\sum_j w_j}$$
where $w_i$ is defined as above for $0 \leq i \leq n$.
\end{thm}
If there is a unique path from every node $j \neq 0$ back to the node $0$, then it follows immediately that there
is exactly one directed spanning tree $T_0$ rooted at $0$ (which represents the open state).  It then follows that the 
steady state probability of the channel being in the open state is of the form
\begin{eqnarray}
\label{eq:MCTOP} \oss &=& \frac{w(T_0)}{\sum_j w_j}.
\end{eqnarray}
As there is only a single term of the form (\ref{eq:prodterm}) in the numerator, it follows readily by combining (\ref{eq:MCTOP}) with 
Lemma 
\ref{lemone} that $\oss$ can will take the form 
\begin{equation}
\oss =\frac{1}{1+\displaystyle\sum_{i=1}^{N}e^{(V-V_{h,i})s_{i}}}.
\label{eq:Msum}
\end{equation}

A few examples of network topologies for which this form is guaranteed by this analysis are illustrated in 
table \ref{tab:tabl1}.

\begin{sidewaystable*}
\begin{tabular}{|m{60mm} m{60mm} m{60mm}|}
\hline
\centering
$\xymatrix{C_{1}\ar@<.5ex>[dr]^{k_{01}} &  & C_{2}\ar@<.5ex>[dl]^{k_{02}}\\
& O\ar@<.5ex>[ul]^{k_{10}}\ar@<.5ex>[ur]^{k_{20}}\ar@<.5ex>[d]^{k_{30}}\\
& C_{3}\ar@<.5ex>[u]^{k_{03}}}$
& 
\centering
$\xymatrix{ & C_{5}\ar@<.5ex>[d]^{k_{15}} &  & C_{6}\ar@<.5ex>[d]^{k_{26}}\\
C_{4}\ar@<.5ex>[r]^{k_{14}} & C_{1}\ar@<.5ex>[l]^{k_{41}}\ar@<.5ex>[u]^{k_{51}}\ar@<.5ex>[dr]^{k_{01}} &  & 
C_{2}\ar@<.5ex>[u]^{k_{62}}\ar@<.5ex>[r]^{k_{72}}\ar@<.5ex>[dl]^{k_{02}} & C_{7}\ar@<.5ex>[l]^{k_{27}}\\
 &  & O\ar@<.5ex>[ul]^{k_{10}}\ar@<.5ex>[ur]^{k_{20}}\ar@<.5ex>[d]^{k_{30}}\\
 &  & C_{3}\ar@<.5ex>[dl]^{k_{93}}\ar@<.5ex>[dr]^{k_{83}}\ar@<.5ex>[u]^{k_{03}}\\
 & C_{9}\ar@<.5ex>[ur]^{k_{39}} &  & C_{8}\ar@<.5ex>[ul]^{k_{38}}}$
 &
 \centering
$\xymatrix{ & O\ar@<.5ex>[dl]^{k_{10}}\ar@<.5ex>[dr]^{k_{20}}\\
C_{1}\ar@<.5ex>[ur]^{k_{01}} &  & C_{2}\ar@<.5ex>[ul]^{k_{02}}\ar@<.5ex>[dl]^{k_{32}}\ar@<.5ex>[dr]^{k_{42}}\\
 & C_{3}\ar@<.5ex>[ur]^{k_{23}}\ar@<.5ex>[dl]^{k_{53}}\ar@<.5ex>[dr]^{k_{63}} &  & C_{4}\ar@<.5ex>[ul]^{k_{24}}\\
C_{5}\ar@<.5ex>[ur]^{k_{35}} &  & C_{6}\ar@<.5ex>[ul]^{k_{36}}}$
\tabularnewline
\centering
Star topology, \newline $\oss=\frac{1}{1+\displaystyle\sum_{i=1}^{3}e^{(V-V_{h,i})s_{i}}}$ & 
\centering
Extended star topology, \newline $\oss=\frac{1}{1+\displaystyle\sum_{i=1}^{9}e^{(V-V_{h,i})s_{i}}}$ &
\centering
Tree topology,\newline $\oss=\frac{1}{1+\displaystyle\sum_{i=1}^{6}e^{(V-V_{h,i})s_{i}}}$
\tabularnewline
\hline 
\centering
$\xymatrix{ & O\ar@<.5ex>[dl]^{k_{10}}\ar@<.5ex>[dr]^{k_{20}}\\
C_{1}\ar@<.5ex>[ur]^{k_{01}}\ar@<.5ex>[rr]^{k_{21}} &  & C_{2}\ar@<.5ex>[ll]^{k_{12}}}$ &
\centering
$\xymatrix{O\ar@<.5ex>[d]^{k_{10}}\ar@<.5ex>[r]^{k_{30}} & C_{3}\ar@<.5ex>[d]^{k_{23}}\\
C_{1}\ar@<.5ex>[u]^{k_{01}}\ar@<.5ex>[r]^{k_{21}} & C_{2}\ar@<.5ex>[u]^{k_{32}}\ar@<.5ex>[l]^{k_{12}}}$ &
\centering
$\xymatrix{O\ar@<.5ex>[d]^{k_{10}}\ar@<.5ex>[r]^{k_{30}} & C_{3}\ar@<.5ex>[d]^{k_{23}}\\
C_{1}\ar@<.5ex>[u]^{k_{01}}\ar@<.5ex>[r]^{k_{21}} & C_{2}\ar@<.5ex>[l]^{k_{12}}}$
\tabularnewline
\centering
$\oss=\frac{1}{1+\displaystyle\sum_{i=1}^{5}e^{(V-V_{h,i})s_{i}}}$ 
& \centering
$\oss=\frac{1}{1+\displaystyle\sum_{i=1}^{9}e^{(V-V_{h,i})s_{i}}}$ & 
\centering
$\oss=\frac{1}{1+\displaystyle\sum_{i=1}^{6}e^{(V-V_{h,i})s_{i}}}$ 
\tabularnewline
\multicolumn{3}{|c|}{Ring topologies with a unique simple path to the open state}
\tabularnewline
\hline
\centering
$\xymatrix{ & O\ar@<.5ex>[dl]^{k_{10}}\ar@<.5ex>[dr]^{k_{20}} &  & C_{3}\ar@<.5ex>[dl]^{k_{23}}\\
C_{1}\ar@<.5ex>[ur]^{k_{01}}\ar@<.5ex>[rr]^{k_{21}} &  & C_{2}\ar@<.5ex>[ll]^{k_{12}}\ar@<.5ex>[ur]^{k_{32}}}$ &
\centering
$\xymatrix{ & O\ar@<.5ex>[dl]^{k_{10}}\ar@<.5ex>[dr]^{k_{20}} &  & C_{3}\ar@<.5ex>[dl]^{k_{23}}\\
C_{1}\ar@<.5ex>[rr]^{k_{21}} &  & C_{2}\ar@<.5ex>[ul]^{k_{02}}\ar@<.5ex>[ll]^{k_{12}}\ar@<.5ex>[ur]^{k_{32}}}$ &
\centering
$\xymatrix{O\ar@<.5ex>[d]^{k_{10}}\ar@<.5ex>[r]^{k_{50}} & C_{5}\ar@<.5ex>[r]^{k_{45}} & C_{4}\ar@<.5ex>[d]^{k_{34}}\ar@<.5ex>[l]^{k_{54}}\\
C_{1}\ar@<.5ex>[u]^{k_{01}}\ar@<.5ex>[r]^{k_{21}} & C_{2}\ar@<.5ex>[l]^{k_{12}}\ar@<.5ex>[r]^{k_{32}} & 
C_{3}\ar@<.5ex>[l]^{k_{23}}\ar@<.5ex>[u]^{k_{43}}}$
\tabularnewline
\centering
$\oss=\frac{1}{1+\displaystyle\sum_{i=1}^{8}e^{(V-V_{h,i})s_{i}}}$ 
& \centering
$\oss=\frac{1}{1+\displaystyle\sum_{i=1}^{7}e^{(V-V_{h,i})s_{i}}}$ & 
\centering
$\oss=\frac{1}{1+\displaystyle\sum_{i=1}^{20}e^{(V-V_{h,i})s_{i}}}$ 
\tabularnewline
\multicolumn{3}{|c|}{Mesh topologies with a  unique simple path to the open state}
\tabularnewline
\hline
\end{tabular}
\caption{Examples of ion-channel topologies that satisfy the criteria of a 
unique simple path to the open state}
\label{tab:tabl1}
\end{sidewaystable*}

\section{Numerical analysis and low order approximation}

We have observed so far that open-state probabilities of ion-channels may be expressed in a similar form to the modified Boltzmann 
equation, 
but with sum of 
exponentials replacing the single exponential term as in equation \eqref{eq:Msum}. The value of $N$ is generally not less than the number 
of transition macro-states. The ambiguity in the value of $N$ (which is at large dependent on the network structure) together with 
computational and identifiability issues for large $N$ motivate us to consider models with small $N$. Of course, when considering such 
approximations, we potentially open up inaccuracies in the model \cite{Liebo89}, and it is therefore important to check that any such 
reduction still accurately captures the observable behaviour of the system.

A good way of approximating the model would be to search for the minimum number 
of exponential terms that yields a good fit for experimentally observed ion 
channel current-voltage characteristic data. Let us denote the $M$-vector of 
ionic currents obtained from patch clamp experiments on a certain channel by 
$\I\in \mathds{R}^M$. Let $\V\in\mathds{R}^M$ be the corresponding voltage vector. 
In order to fit these data to equation \eqref{eq:Msum}, the distance between the 
experimental data and the model needs to be minimized. The distance may be measured as square of the $L_2$ norm:
\[J(\I,\Ie(\V)) = \|\I-\Ie(\V)\|_2^2 =\sum_{j=1}^m\left(I^m_j-I(V^m_j)\right)^2\]
where $\Ie(\V)\in\mathds{R}^M$ is the vector of ionic current predicted by the 
model given membrane potentials in the vector $\V$. $I^m_j$ and $V^m_j$ are respectively the $j^{th}$ position 
in vectors $\I$ and $\V$. Let $g_{max}$ be the maximal conductance of the channel 
and $V^*$, the equilibrium potential of the ion transported by the channel. 
Mathematically, this optimization problem can be formulated as following:
\begin{equation}
\min_{\{V_{h,i}\}_{i=1}^N,\{s_i\}_{i=1}^N,g_{max}} J(\I,\Ie(\V))
\end{equation}
subject to the conductance expression:
\begin{equation}
I(V_j)=\frac{g_{max}(V_j-V^*)}{1+\displaystyle\sum_{i=1}^N e^{(V-V_{h,i})s_{i}}} 
\quad \forall j=1,..,m
\end{equation}
and with  bounds that are physiologically justified.

This optimization problem brings forth two significant difficulties: (i) the cost 
function is  not convex and (ii) the number of dominant conformations that the 
protein adopts is usually unknown and hence the order of the exponentials, $N$ 
is unknown. To guarantee that the global optimum is achieved, an initial guess should be carefully selected. It is possible to have a 
calculated guess for 
the $V_h$ values from the current-voltage characteristics, positioned centrally 
over the range of values at which the curve shows steady activation or inactivation. The steepness of the tangent drawn at this estimated 
voltage may be used as an initial estimate of the slope value. Alternatively, the recently developed global optimization Scatter Search 
based methodology, SSm~GO \cite{egea2010evolutionary}, can be used. This algorithm combines a population-based metaheuristic 
method with a local optimization.

Regarding the number of exponentials, the aim is to curtail the number of conformational transitions so that it leads to minimum parameters 
to fit the data. For this purpose an algorithm was implemented which calculates the global optimum for several order of exponentials until 
the confidence in the fit is accomplished. The algorithm starts by $N=1$ and progressively increases until  the best fit between the model 
and the data has a relative error less than $\epsilon$ times the original data. The measure considered is again the square of the $L_2$ 
norm 
and mathematically, this criterion may be formulated as: 
\[J(\I,\Ie(\V))< \|\epsilon \I\|_2^2\]
For all the cases we have considered, a good fit is obtained by setting 
$\epsilon=0.10$.

Interestingly, for all these cases we have examined, a maximum of two exponentials, i.e. $N=2$ was
required. The data could be accurately fitted by using a two dimensional state space ($N=1$) for
simple symmetric data-plots, and for all other cases, a three dimensional state space ($N=2$) could
accurately explain the data. This would imply that to a large extent stationary distributions of
voltage gated ion-channels can be adequately modeled as dwelling predominantly among three different
macro-states even if they are able to change between several conformations. The time that the
protein dwells in some of these states may be considerably smaller than that of the three dominant
conformations. The three macrostates may incorporate some of the effects of the minor conformations
rather than completely negating their influences; as reflected from the data-fits. 

With the above argument we propose the following equation for the open state probability corresponding to a three state system,
\begin{equation}
\oss=\frac{1}{1+e^{\left[(V-V_{h_1})s_1\right]}+e^{\left[(V-V_{h_2})s_2\right]}}
\label{eq:middleton}
\end{equation}
as a good approximation to represent the stationary probability of voltage gated ion-channels to remain open. The channel current may hence 
be calculated with the following equation:

\begin{equation}
I_{i}=\frac{g_{max,i} (V-V_{i}^{*})}{1+ e^{\{V-V_{h,1}(i)\}s_1(i)} + 
e^{\{V-V_{h,2}(i)\}s_2(i)}}
\label{eq:current}
\end{equation}
where, $I_{i}$ is the channel current; $V_{i}^{*}$, the reversal potential of ion 
$i$; $V$, the membrane potential; $g_{max,i}$, the maximal conductance of the channel 
for the ion $i$ and ${V_{h,1}(i)}, {s_1}(i), {V_{h,2}(i)}, {s_2}(i)$ being the 
parameters for the activation function as defined by equation \eqref{eq:middleton}.

\section{Fitting of experimental data to the model}

\begin{figure*}[!ht]
\begin{centering}
\subfigure[Normalized current-voltage relationship of the T-type calcium channel
from the digitized data of Talavera and Nilius (2006) \cite{Talavera2006}]
{
\includegraphics[width=0.45\textwidth]{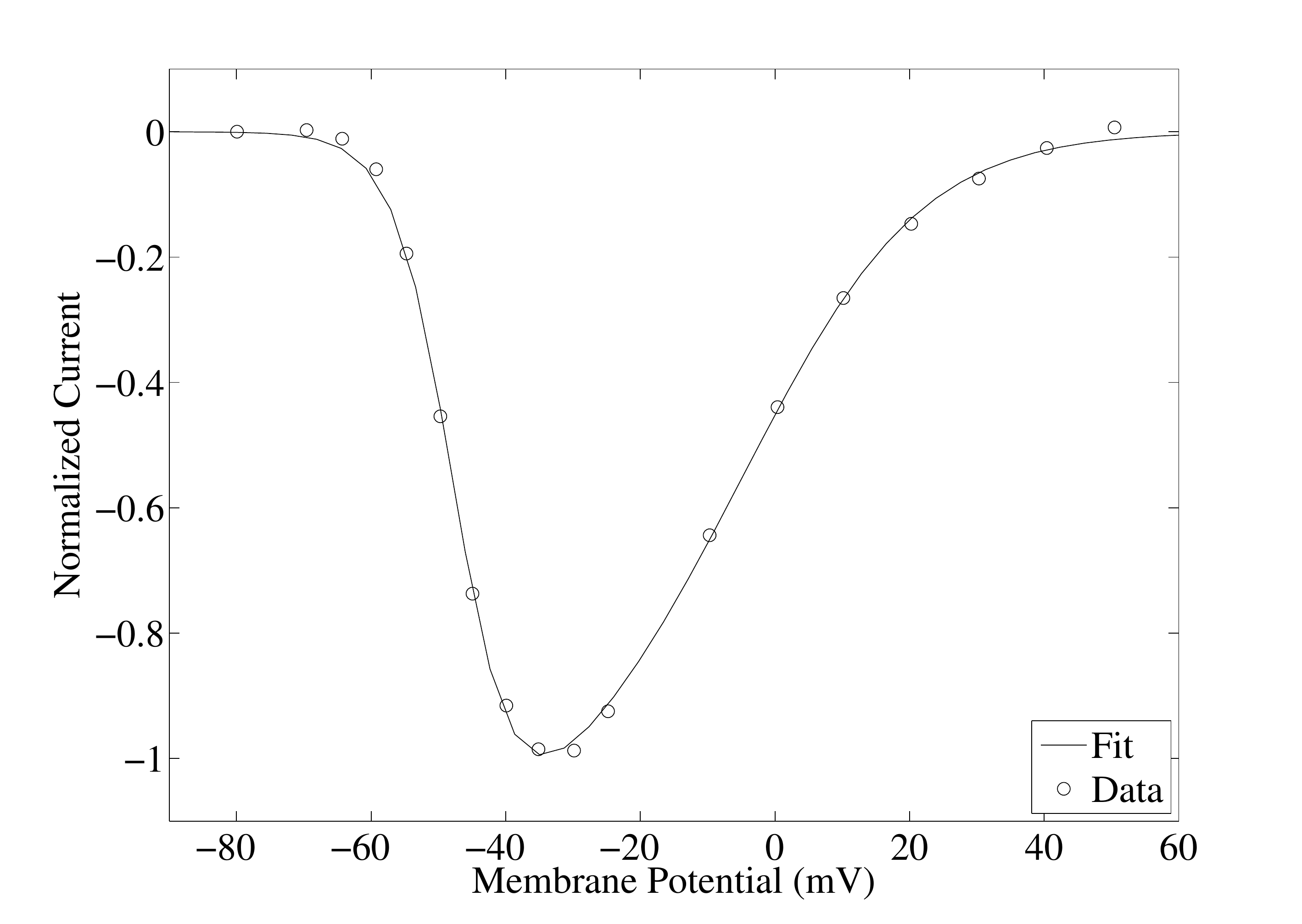}
\label{fig:CaA}
}
\hfill
\subfigure[Normalized current-voltage relationship of the wild-type calcium channel \cite{Beyl2007}]
{
\includegraphics[width=0.45\textwidth]{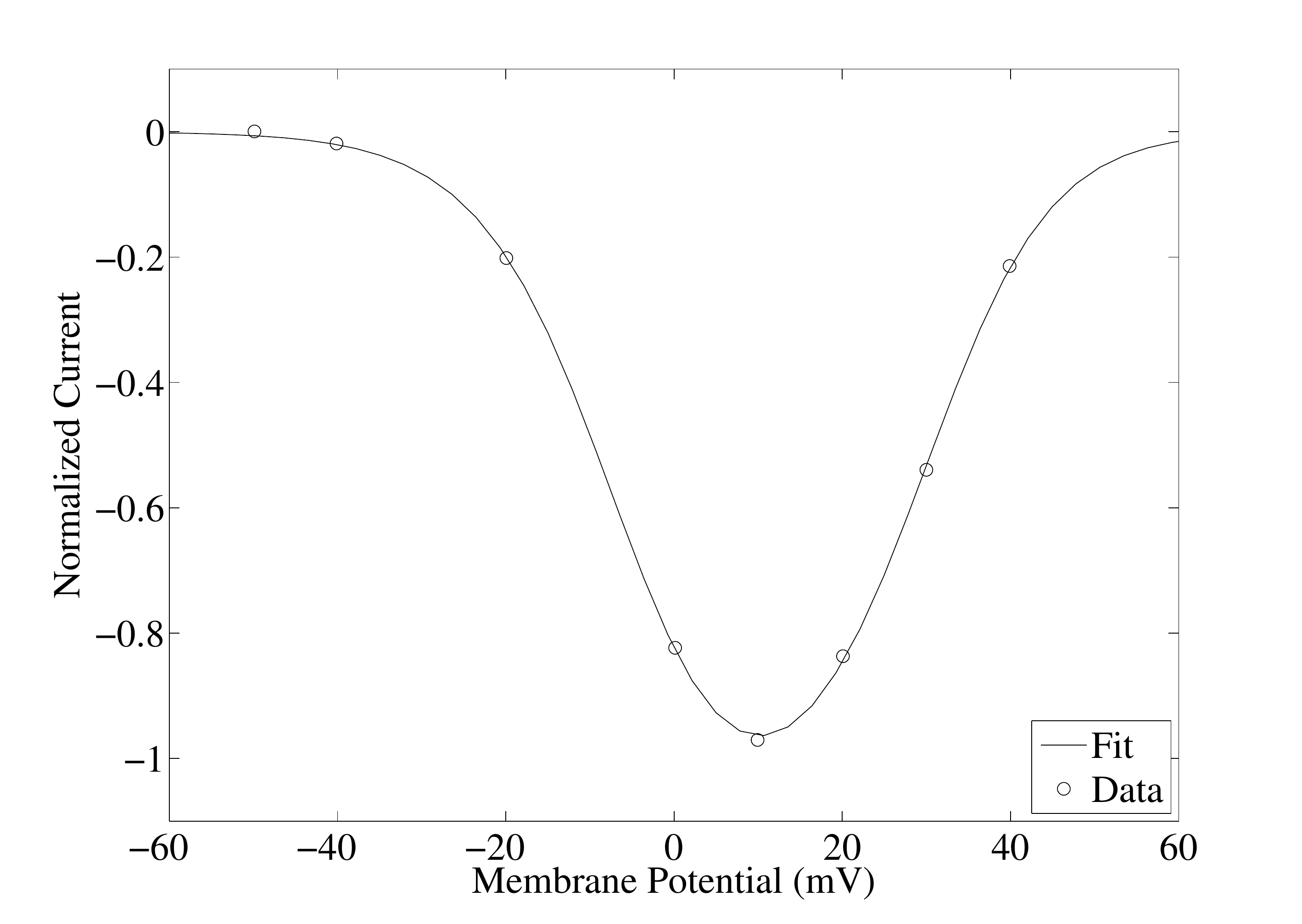}
\label{fig:CaB}
}
\newline
\subfigure[Averaged, peak current\textendash{}voltage plots of $Ca_{v}1.2\alpha$ L-type 
calcium channels \cite{Xu2001}]
{
\includegraphics[width=0.45\textwidth]{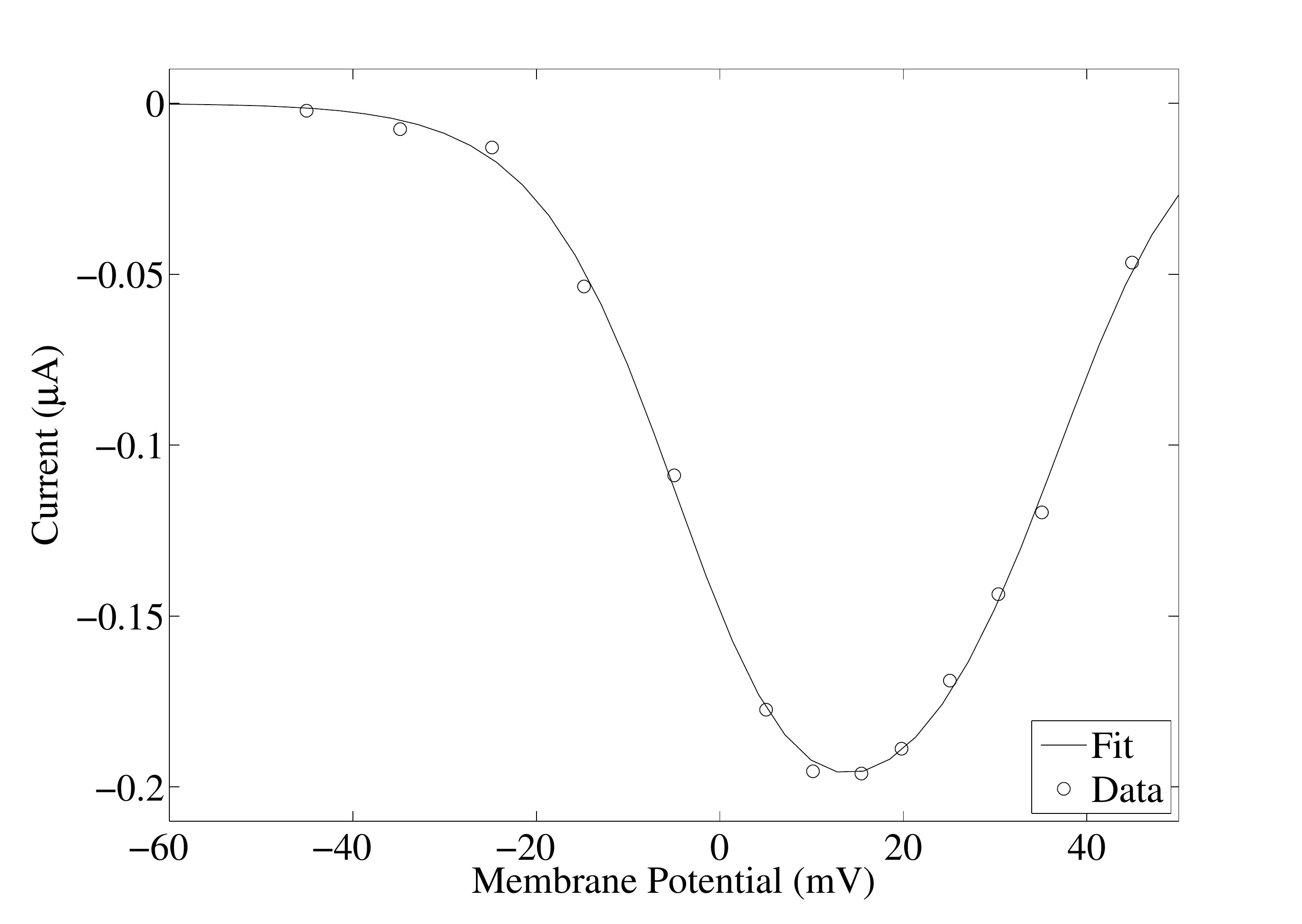}
\label{fig:CaC}
}
\hfill
\subfigure[Averaged, peak current\textendash{}voltage plots of $Ca_{v}1.3\alpha$ L-type 
calcium channels \cite{Xu2001}]
{
\includegraphics[width=0.45\textwidth]{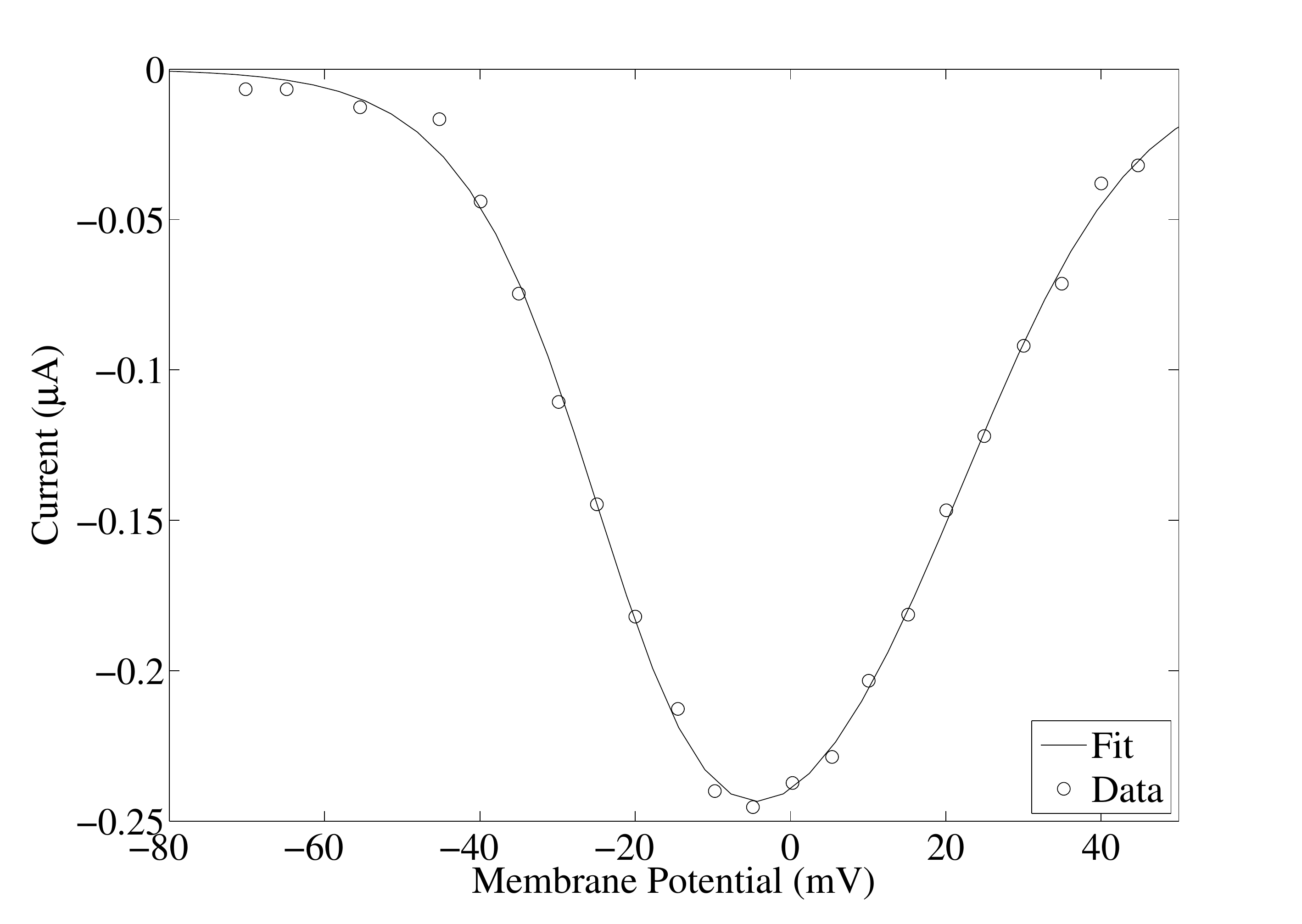}
\label{fig:CaD}
}
\end{centering}
\caption{Current-Volatge relationship of Voltage-gated $Ca^{2+}$ channels fitted with 
equation \eqref{eq:middleton}. Voltage-gated $Ca^{2+}$ channels (VGCC) can be classified 
into high-voltage-activated (HVA) and low-voltage-activated (LVA) channels, which 
implies that LVA channels activate at 20 to 30 mV more negative potentials than HVA channels. T-type calcium channels are LVA and show fast 
macroscopic inactivation where as L-Type calcium channels are HVAs.}
\label{fig:cal}
\end{figure*}

\begin{figure*}[!ht]

\begin{centering}

\subfigure[Peak inward current vs. voltage relation for the current records of
sodium channel \cite{Stuehmer} fitted with equation \eqref{eq:middleton}]
{
\includegraphics[width=0.45\textwidth]{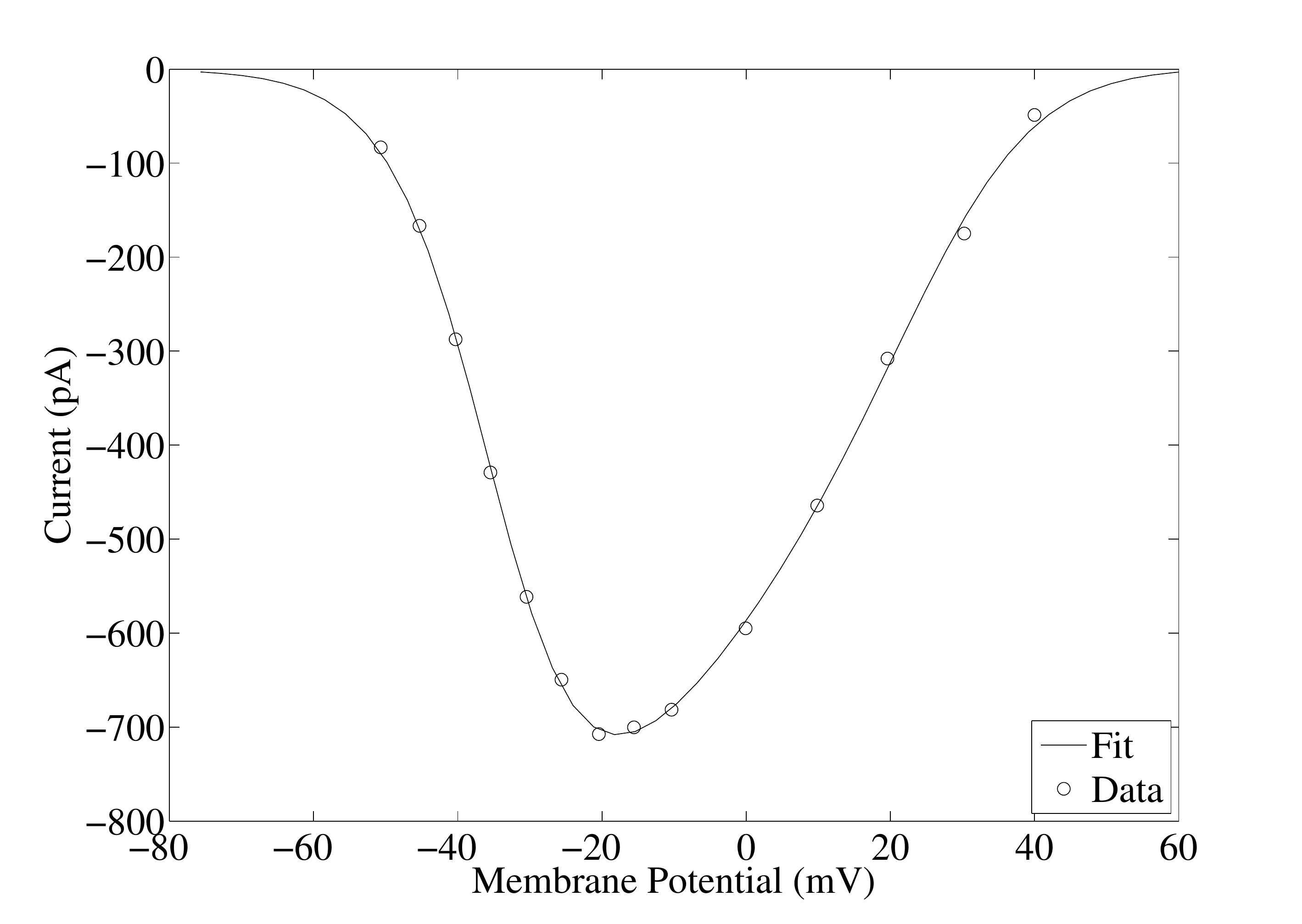}
\label{fig:NaA}
}
\hfill
\subfigure[Normalized current-voltage relationship of sodium channel from the
macropatch current \cite{Ruben1997} fitted with equation
\eqref{eq:middleton}]
{
\includegraphics[width=0.45\textwidth]{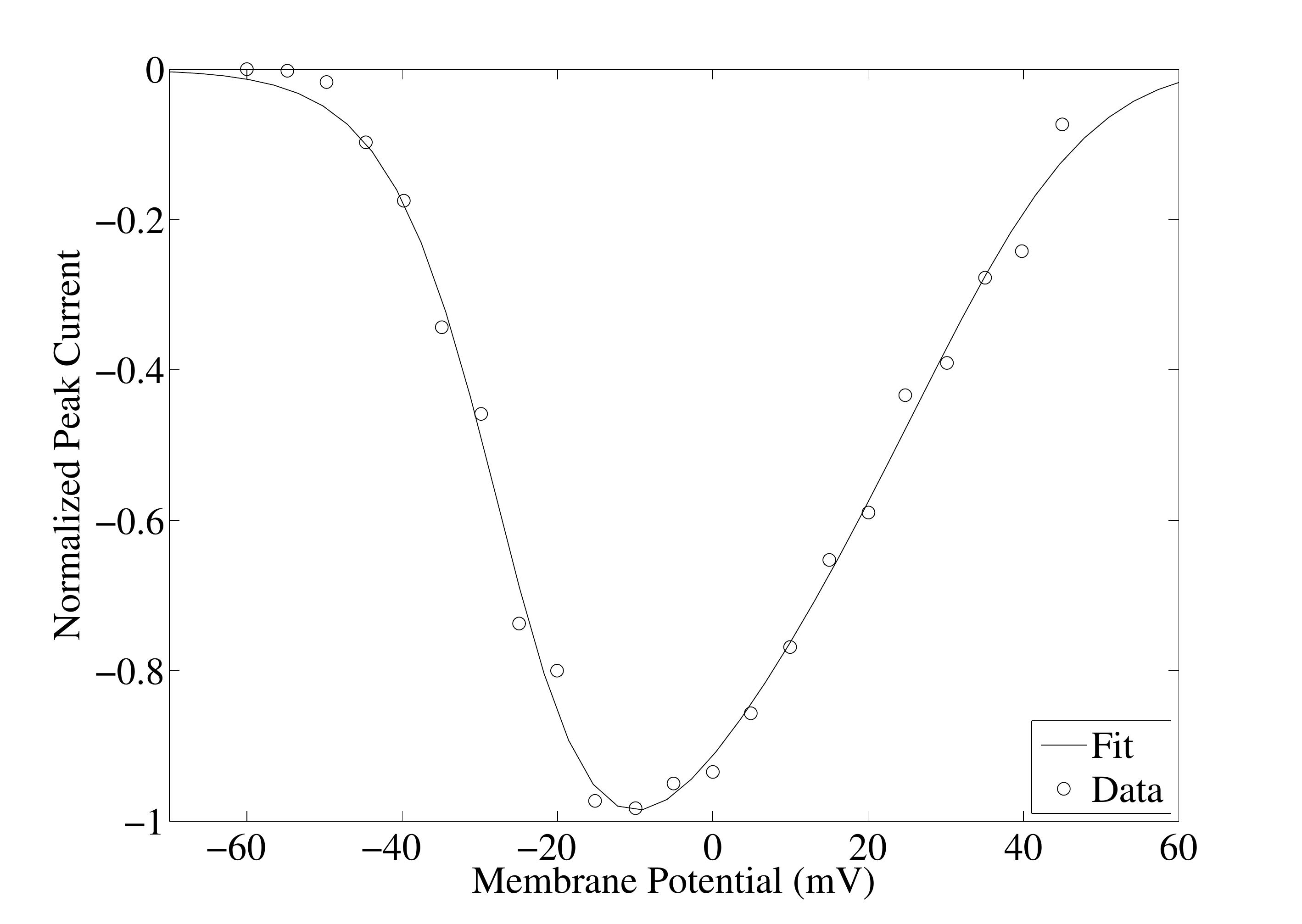}
\label{fig:NaB}
}
\end{centering}
\caption{Current-Volatge relationship of Voltage-gated $Na^+$ channels fitted with 
equation \ref{eq:middleton}.
}
\label{fig:sod}
\vspace{1cm}
\end{figure*}

\begin{figure*}[ht]
\begin{centering}
\subfigure[KCNH5]
{
\includegraphics[width=0.45\textwidth]{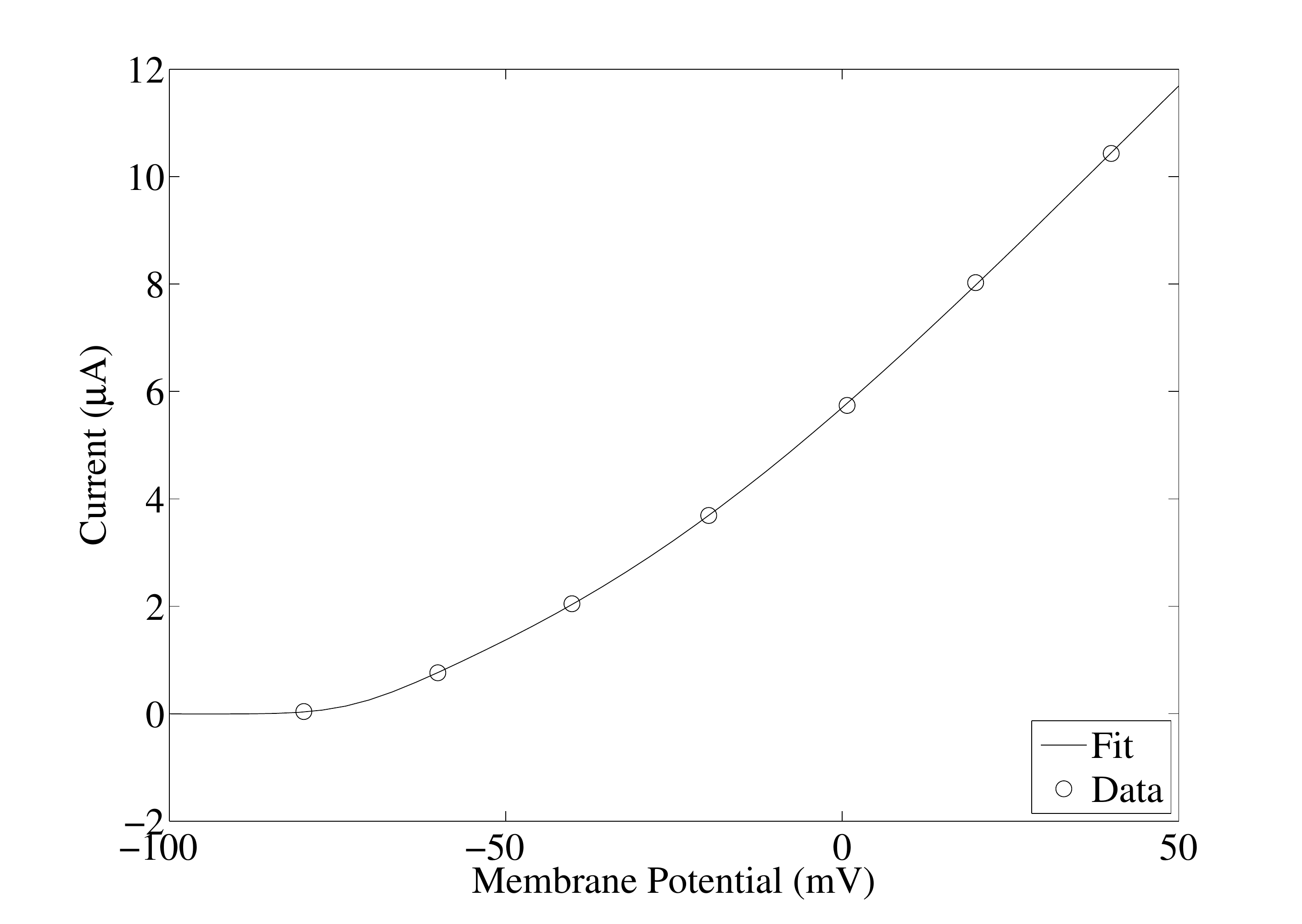}
\label{fig:Ka}
}
\hfill
\subfigure[KCNH7]
{
\includegraphics[width=0.45\textwidth]{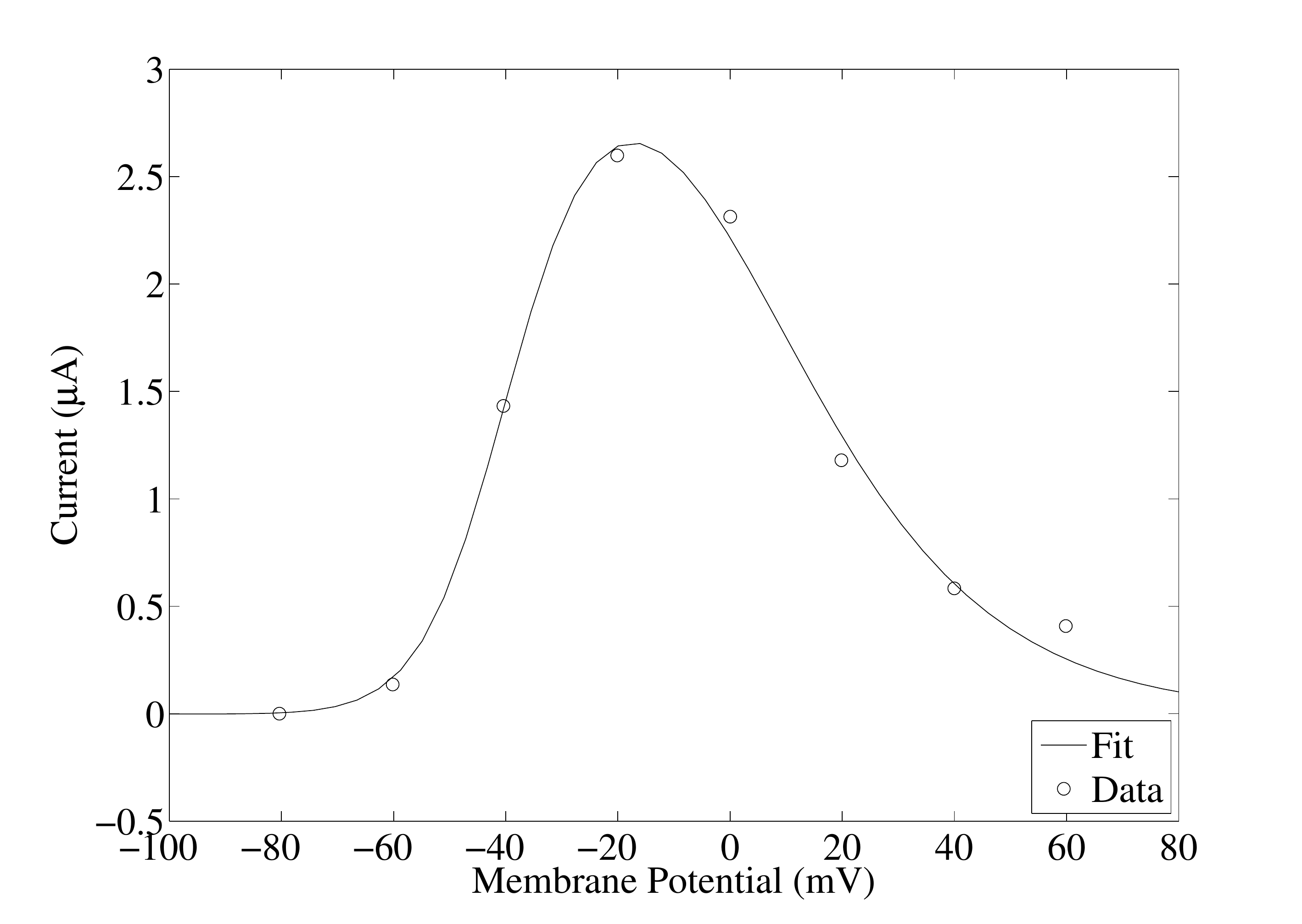}
\label{fig:Kb}
}
\newline
\vspace{-0.5cm}
\begin{center}
\subfigure[KCNB1]
{
\includegraphics[width=0.45\textwidth]{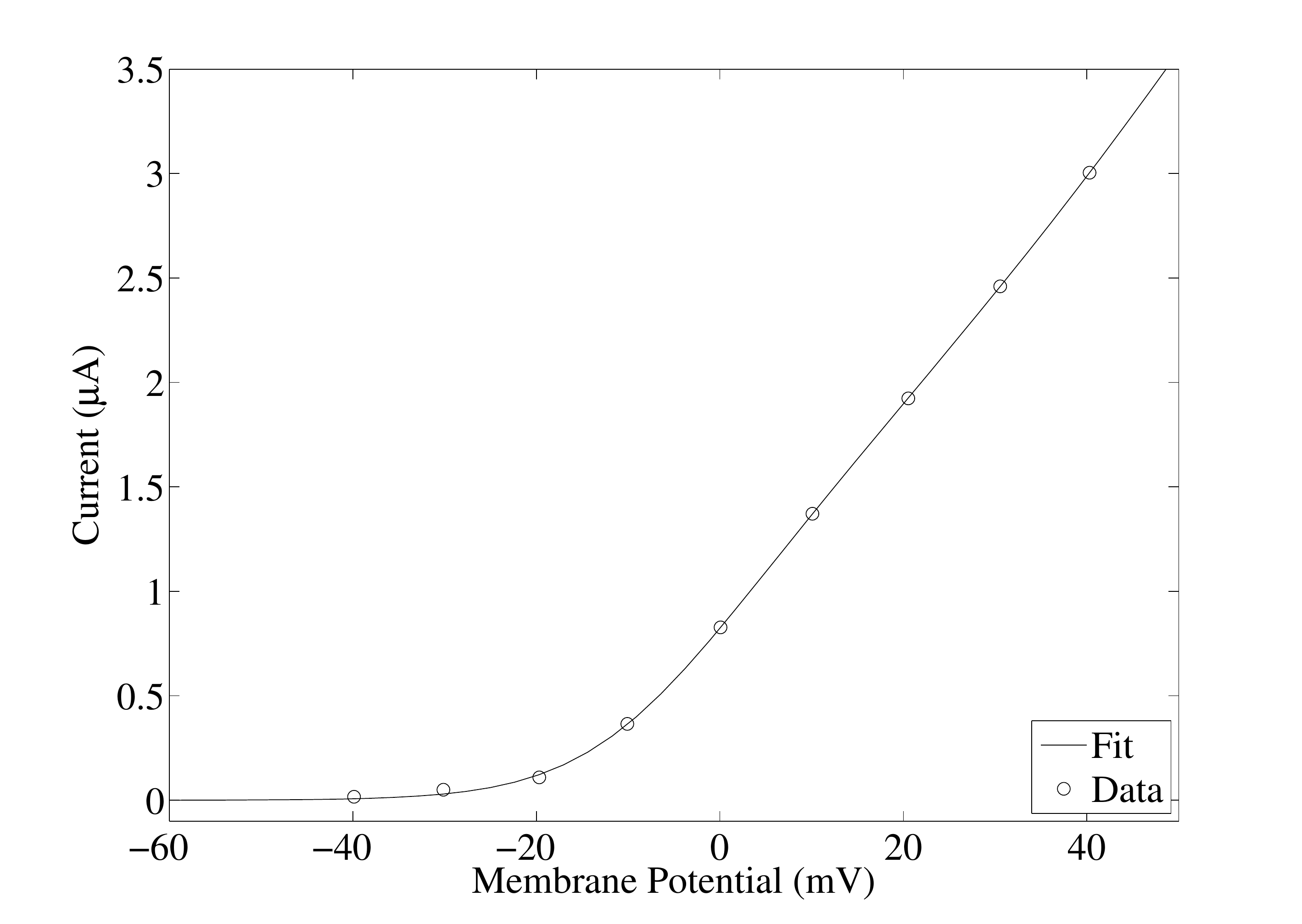}
\label{fig:Kc}
}
\end{center}
\vspace{-0.5cm}
\end{centering}
\caption{Current-voltage relations for KCNH5, KCNH7 
and KCNB1 class of potassium channels expressed in oocytes \cite{Zou2003} fitted with equation 
\ref{eq:middleton}.
}
\label{fig:pott}
\end{figure*}

Most papers in the literature follows the voltage step protocol where steady-state voltage-current characteristics are obtained 
independently of the ion channel dynamics. Several were selected for utilizable data on single channel studies. We collect data from both, 
stationary conductance obtained from peak currents (with less signal-to-noise ratio \cite{beaumont1993estimation}) or from raw 
steady-state measurements. Data were extracted from the published curves using the Enguage Digitizer 4.1. The digitized data sets were fit 
by the equation \eqref{eq:current}. The global scatter search algorithm, SSm GO \cite{egea2010evolutionary} 
implemented in MATLAB was used for making the fits. The resulting data fits are presented in figures 
\ref{fig:cal},\ref{fig:sod},\ref{fig:pott} and summarized in table \ref{tab:parfit}. In all cases, with $N \le 2$, we are able to represent 
the observed data well using the model structure proposed.


\begin{table*}[ht]

\centering{}\begin{tabular}{c|c|c|c|c|c|c|c}

Figure  & Reference  & Channel type  & $V_{h_{1}}$ ($mV$)  & $s_{1}$ ($mV$)  & $V_{h_{2}}$ ($mV$)  & $s_{2}$ ($mV$)  & $g_{max}$ \\
\hline
\ref{fig:CaA}  & Talavera and Nilius (2006) \cite{Talavera2006}  & $Ca_{v}3.1$  & -47.2228  & -0.22613  & 0.617753  & 0.07519  & 0.0069 
($mV^{-1}$)  \\

\ref{fig:CaB}  & Beyl et al. (2007) \cite{Beyl2007}  & $Ca_{v}1.2$  & -5.36225  & -0.12598  & 31.7746  & 0.13336  & 0.0098 ($mV^{-1}$)  \\

\ref{fig:CaC}  & Xu and Lipscombe (2001) \cite{Xu2001}  & $Ca_{v}1.2$  & -2.89138  & -0.13021  & 38.9031  & 0.13943  & 0.002 ($\mu A/mV$)  
\\

\ref{fig:CaD} & Xu and Lipscombe (2001) \cite{Xu2001}  & $Ca_{v}1.3$  & -22.4361  & -0.11182  & 26.4652  & 0.08824  & 0.0022 ($\mu A/mV$)  
\\

\ref{fig:NaA} & Stühmer et al. (1987) \cite{Stuehmer}  & $Na_{v}1.2$  & -34.1391  & -0.14523  & 28.3072  & 0.10388  & 8.8843 ($pA/mV$)  \\
 
\ref{fig:NaB} & Ruben et al. (1997) \cite{Ruben1997}  & $Na_{v}1.2a$  & -25.6539  & -0.14232  & 38.2236  & 0.088449  & 0.0138 ($mV^{-1}$)  
\\

\ref{fig:Ka}  & Zou et al. (2003) \cite{Zou2003}  & $K_{v}10.2$  & -64.5494  & -0.17538  & -29.4952  & -0.02195  & 0.1009 ($\mu A/mV$)  \\

\ref{fig:Kb} & Zou et al. (2003) \cite{Zou2003}  & $K_{v}11.3$  & -41.6897  & -0.11727  & -6.79872  & 0.053344  & 0.0630 ($\mu A/mV$)  \\

\ref{fig:Kc} & Zou et al. (2003) \cite{Zou2003}  & $K_{v}2.1$  & 40  & -0.021769  & 3.51185  & -0.13161  & 0.0476 ($\mu A/mV$) \\

\end{tabular}
\caption{\label{tab:parfit}Fitted parameters of Channel current - membrane potential data of a few ion-channels using equation 
\ref{eq:middleton} }
\end{table*}

\section{Comparison with non-linear thermodynamic models}
\label{sec:nl}

\begin{figure*}[!ht]
\begin{centering}
\subfigure[Calcium channel\cite{Talavera2006}]
{
\includegraphics[width=0.45\textwidth]{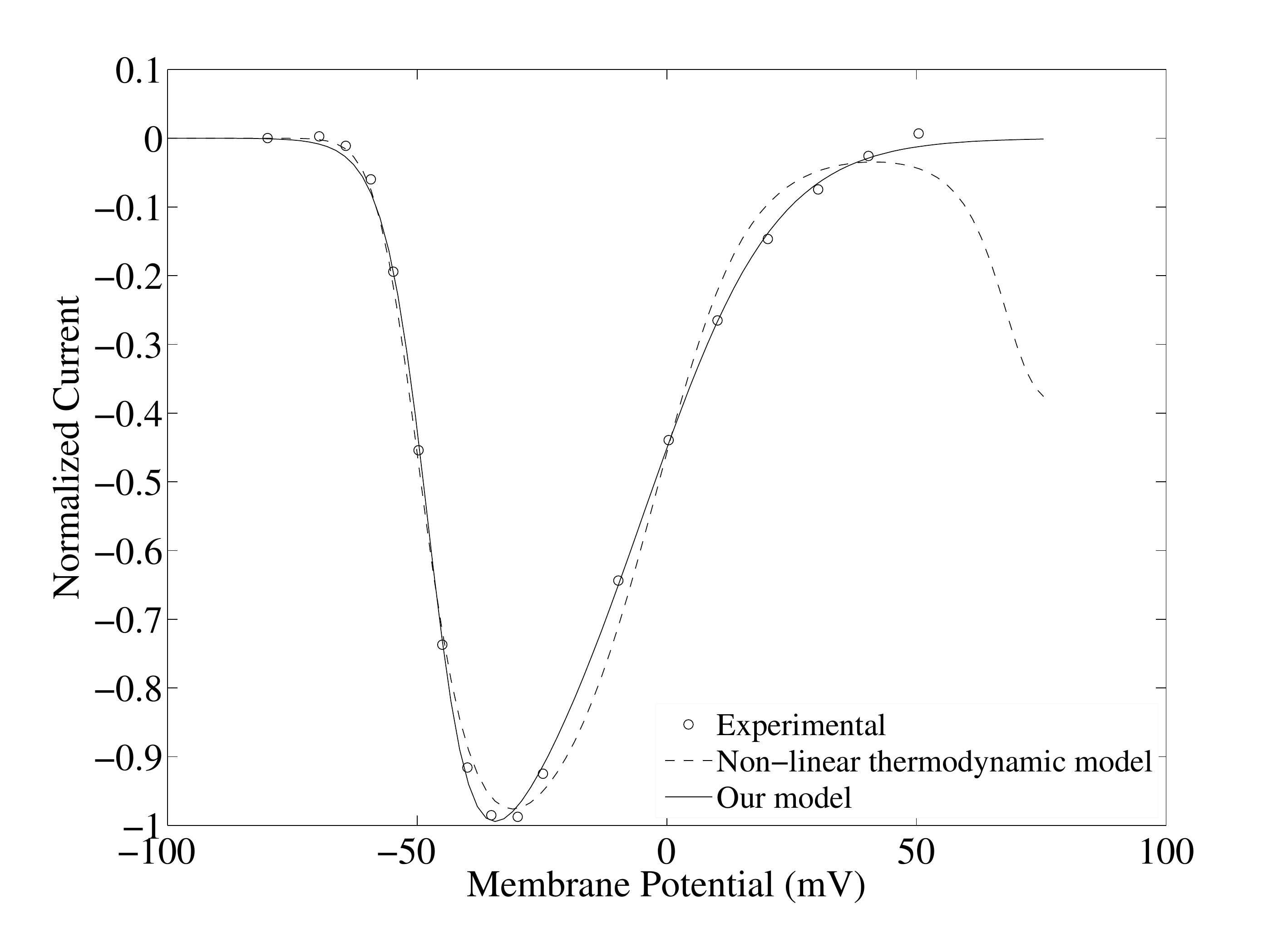}
\label{fig:thermA}
}
\hfill
\subfigure[Potassium channel\cite{Zou2003}]
{
\includegraphics[width=0.45\textwidth]{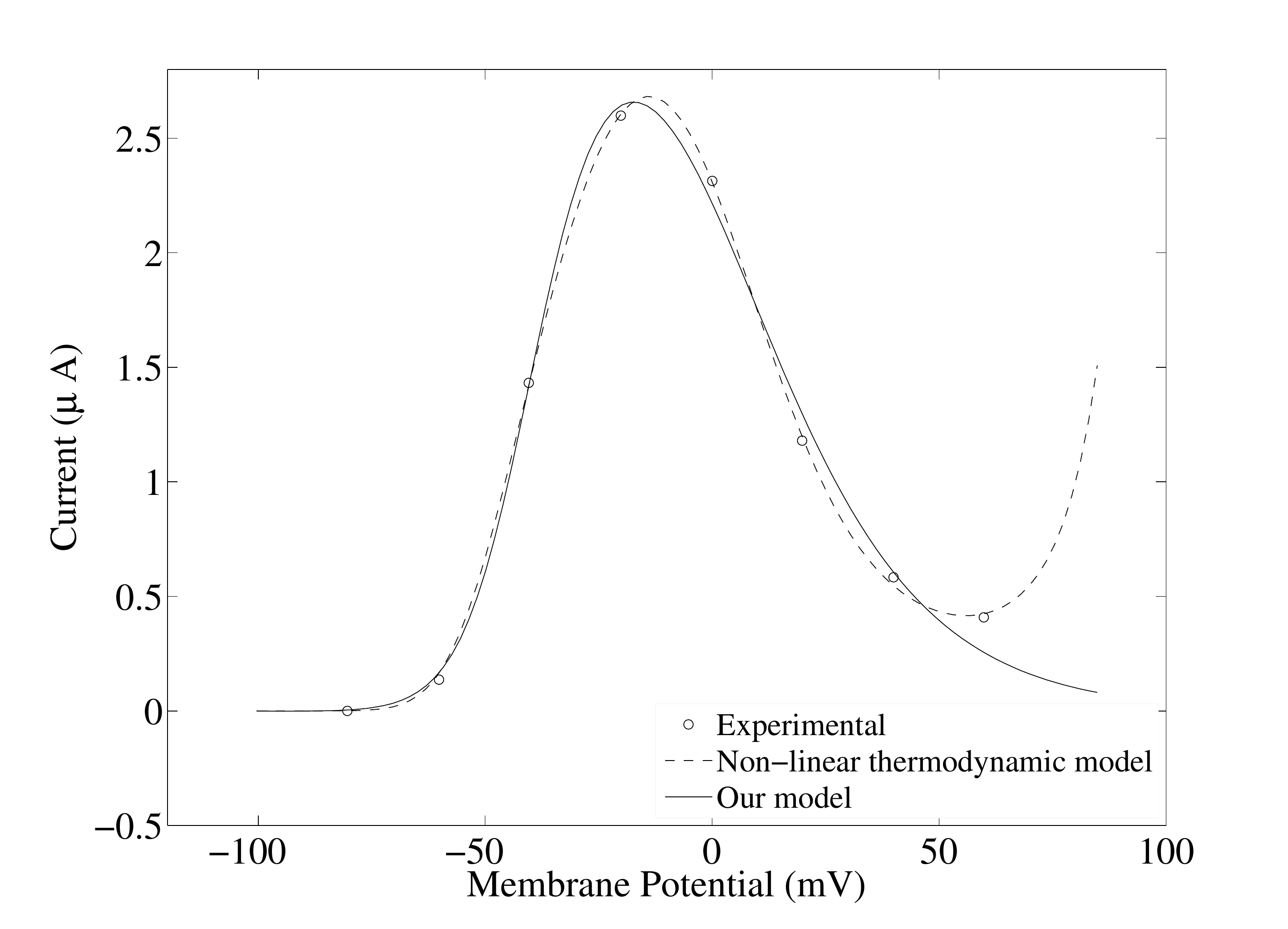}
\label{fig:thermB}
}
\end{centering}
\caption{Current-voltage relationship of ion-channels obtained from the studies \cite{Talavera2006,Zou2003} fitted with 
a third order non-linear model and modified model presented in this article according to equation \eqref{eq:middleton}}
\label{fig:therm}
\end{figure*}

The estimation of the open-state probability of voltage-gated ion channels to an equation of form as (\ref{eq:Msum}) would rather be 
equivalent to a steady-state approximation of the Hodgkin-Huxley formalism. However, as we seek to look beyond the Hodgkin-Huxley formalism 
for a good mechanistic description, the given model need to be compared with efforts made to represent the system on a realistic 
perspective. Thermodynamic formalism comes closest to this effort. Although the fundamentals are in place, such models \cite{Destexhe2000} 
show very poor performance in even small extrapolations from the given data (Figure \ref{fig:therm}). We would also argue that the 
biophysical basis of the multiple conformation model is much clearer than that of a Taylor's series dual conformation model.

Ozer \cite{Ozer2004,Ozer2007} modified the non-linear model to a functional form by lumping the different transitions in the protein to a 
single event. The model which uses a sum of Gaussion distribution, seems to give the best fits for experimental data. The steady-state open 
probability of ion-channels written with respect to this model,
may be arrived using equation (8) of \cite{Ozer2007} as,
\begin{equation}
\oss = \frac{1}{1+\frac{\displaystyle \sum_{i=1}^n \alpha_{0,i} e^{[(V-V_{\alpha,i})s_{\alpha,i}]^2}}
{\displaystyle \sum_{i=1}^n \beta_{0,i} e^{[(V-V_{\beta,i})s_{\beta,i}]^2}}}
\label{eq:ozer}
\end{equation}
where, $n$ was defined as the number of distinct transitions defined by different energy 
barriers. It should be noted that in work \cite{Ozer2007} was also able to get acceptable fits with two transitions. The problem with this 
model 
however seems to be in the existence of unidentifiable parameters, as is the case with majority of the existing Markov Models 
\cite{Fink2009}. Further, if no manipulation is carried out to collect group of parameters that are no identifiable only with stationary 
data, the model asks for fitting a minimum of twelve parameters (provided two macro-transitions ($n = 2$) gives acceptable fits).

A mathematical model for a biological phenomenon is assured to have uniquely estimated parameters if the model structure ensures 
identifiability \cite{walter1997}. Conventionally, ion-channel data has limited interpretations by producing indistinguishable models 
\cite{Kienker1989}. In-silico representation of ion-channel dynamics is yet to come up with a model for the reason that the model 
structure needs to be tweaked each time to incorporate experimental observation. This indeed is a question of parameter identifiability.  
The model that we have presented in this article, is found to have \emph{structurally output locally identifiable (s.o.l.i)} parameters 
(see \ref{app:c} for formal definition), according to the following lemma. 

\begin{lem}
\label{lem:identifiability}
The model for ion channel open state probability described by means of a Markov chain with three macro-states [equation 
(\ref{eq:middleton})] is s.o.l.i by the Taylor approximation of fourth order.
\end{lem}
\emph{Proof:} see \ref{app:c}

\section{Conclusion}

In this article, we bring forth an important criterion in modelling ion channel activity using Markov jump schemes.
We propose that the behaviour of the system may be developed by an analytical design that is simple and adaptable; and towards this, a 
minimum of three Markov states are necessary to model the dynamics.

To illustrate the criterion, we describe the physical transitions of a particular class of voltage gated ion channel proteins by means  of 
free energy changes associated with perturbations in its environmental electric field from a thermodynamic perspective. 
For this class of ion-channels we assume the existence of a unique simple path of transition, from every stable state to the open state.
The resulting model is a simple kinetic expression (equation \ref{eq:middleton}) which approximates the stationary open probability of the 
channel for a given membrane potential and is a generalized form of the single exponent modified Boltzmann's function.

The model proposed in this article, is identifiable and requires small number of parameters compared to existing models. This would in turn 
reduce the computational cost associated with regular Markov Models, without compromising much from the mechanistic viewpoint. Its main 
limitation is that it has been developed for steady-state data and may not represent ion-channel dynamics in its entirety but only fast ion 
channels. Nonetheless, when dynamics are relevant, the proposed scheme may be extended and the resulting model calibrated by using 
voltage-current data from classical experimental protocols.


%
\section*{Acknowledgment}

The authors are grateful to the Science Foundation of Ireland for funding the research (Science Foundation of Ireland Research Grant 
07/PI/I1838). We are also indebted to Professor Peter Wellstead and Wilhelm Huisinga for their valuable support and helpful discussions.

\bibliographystyle{plain}

\appendix
\section{Model Identifiability, Proof of Lemma \ref{lem:identifiability} }
\label{app:c}
Simplicity and identifiability of ion channel model would be of advantage especially, when such models are employed to develop larger 
frameworks describing metabolic activity of a cellular system, such a pacemaking neurons \cite{francis2013single}
or cardiac cells \cite{Matsuoka2007}. To study the parameter identifiability of model given by equation (\ref{eq:middleton}), let us 
consider the following general mathematical structure:
\[M(\theta):y(\theta)=f(\theta,x)\]
where $\theta\in\mathds{R}^{n_\theta}$ is the set of parameters to be estimated and $y(\theta)\in\mathds{R}$ and $x\in\mathds{R}$ denote 
the 
output and input of the system, respectively. In addition, let us denote the search space of the parameter by 
$\Theta\subseteq\mathds{R}^{n_\theta}$.

\begin{mydef}\label{def:identifiability}

The model $M(\theta)$ is said to be \textbf{structurally output globally identifiable} (s.o.g.i) , if for any 
$\widetilde{\theta}\in\Theta$, 
except for the points of a subset of measure zero, and for all $x\in\mathds{R}$:
\[y(\theta,x)\equiv y(\widetilde{\theta},x)  \Longrightarrow \theta \equiv \widetilde{\theta}\]
If the same conditions is fulfilled only in the neighborhood of $\theta$ then the parameters will be \textbf{structurally output locally 
identifiable} (s.o.l.i.) 
\end{mydef}

This definition is quite general and is often difficult to be put into practice. One of the usual approach in such instance is the 
so-called 
"Taylor approach" \cite{walter1997} where both outputs $y(\theta,x)$ and $y(\widehat{\theta},x)$ are approximated by a Taylor expansion 
around a point $x^*\in \mathds{R}$. If we denote by $\mathcal{U}_{x^*}^n$ the ball around $x^*$, where the Taylor approximation of order 
$n$ 
holds; the general identifiability definition is therefore set as:
\begin{mydef}
The model $M(\theta)$ is said to be s.o.g.i in $x\in\mathcal{U}_{x^*}^n$ if the system of equations:
\begin{equation}\label{eq:coefficients}
\left[\frac{\partial^i y(\theta,x)}{\partial x^i}\right]_{x=x_0}=\left[\frac{\partial^i y(\widetilde{\theta},x)}{\partial 
x^i}\right]_{x=x_0} \quad \forall i=1,...n
\end{equation}
has an unique solution. In the same way, the model will be s.o.l.i in $x\in\mathcal{U}_{x^*}^n$ if a finite number of solution is obtained.
\end{mydef}

\begin{proof}

It has to be proved that, the ion channel open state probability represented by the equation,
\begin{equation}\label{eq:to_identi}
G(\theta) = \frac{1}{1+e^{(V-V_{h_1})s_1}+e^{(V-V_{h_2})s_2}}
\end{equation} 
is s.o.l.i for any $V\in\mathcal{U}_{V^*}^n$, by using the Taylor approximation of fourth order. It may be noted that studying the 
identifiability property for $G(\theta)$ with parameters
\[\theta=[V_{h,1},s_1,V_{h,2},s_2]\in\mathds{R}^4 \qquad 
\widetilde{\theta}=[\widetilde{V}_{h,1},\widetilde{s}_1,\widetilde{V}_{h,2},\widetilde{s}_1]\in\mathds{R}^4\]
 is equivalent to studying this property for the model
\[M(\theta) = e^{(aV+b)}+e^{(cV+d)}, \quad \theta = [a,b,c,d]\]
where
\[a = s_1, \quad b = -V_{h_1}s_1, \quad c =s_2, \quad d =-V_{h_2}s_2 \]
\[\at =\widetilde{s}_1, \quad \bt =-\widetilde{V}_{h_1}\widetilde{s}_1, \quad \ct =\widetilde{s}_2, \quad \dt 
=-\widetilde{V}_{h_2}\widetilde{s}_2 \]
Therefore, the system of equations built by using (\ref{eq:coefficients}) leads to:
\begin{subequations}\label{eq:sys_eq}
\begin{equation}
e^{aV_0+b}+e^{cV_0+d}=e^{\at V_0+\bt}+e^{\ct V_0+\dt}
\end{equation}
\begin{equation}
ae^{aV_0+b}+ce^{cV_0+d}=\at e^{\at V_0+\bt}+\ct e^{\ct V_0+\dt}
\end{equation}
\begin{equation}
a^2e^{aV_0+b}+c^2e^{cV_0+d}=\at^2 e^{\at V_0+\bt}+\ct^2 e^{\ct V_0+\dt}
\end{equation}
\begin{equation}
a^3e^{aV_0+b}+c^3e^{cV_0+d}=\at^3 e^{\at V_0+\bt}+\ct^3 e^{\ct V_0+\dt}
\end{equation}
\end{subequations}
with the following two solutions obtained with Mathematica software \cite{mathematica}:
\[a = \ct, \quad b = \dt, \quad c = \at, \quad d=\bt\]
\[a = \at, \quad b = \bt, \quad c = \ct, \quad d=\dt\]
Therefore, the system is s.o.l.i in the neighbourhood of $V^*$ where the Taylor approximation holds.
\end{proof}

\begin{cor}
 If the model (\ref{eq:to_identi}) is completed with the following condition:
 \[s_1>s_2\]
it is trivial to see that the system (\ref{eq:sys_eq})now has only one solution and the model is s.o.g.i. for any $V\in\mathcal{U}_{V^*}^n$.
\end{cor}



\end{document}